\pdfoutput=1
\documentclass[11pt,a4paper,american]{extarticle}
\usepackage{a4wide}
\usepackage[active]{srcltx}
\usepackage[export]{adjustbox}
\usepackage{amssymb}
\usepackage{amsmath}
\usepackage{amsfonts}
\usepackage{amsthm}
\usepackage{array,multirow,bigdelim}
\usepackage{authblk}
\usepackage{babel}
\usepackage{bbm}
\usepackage{bm}
\usepackage{braket}
\usepackage{cite}
\usepackage{color}
\usepackage{comment}
\usepackage{enumerate}
\usepackage{esint}
\usepackage{etoolbox}
\usepackage{float}
\usepackage[T1]{fontenc}
\usepackage{graphicx}
\usepackage[latin1]{inputenc}
\usepackage{latexsym}
\usepackage{mathrsfs}
\usepackage{mathtools}
\usepackage{relsize}
\usepackage{setspace}
\usepackage{slashed}
\usepackage{soul}
\usepackage{subcaption}
\usepackage{todonotes}
\usepackage{xcolor}
\usepackage{hyperref}
\newcommand{\xdownarrow}[1]{%
	{\left\downarrow\vbox to #1{}\right.\kern-\nulldelimiterspace}
}

\apptocmd{\sloppy}{\hbadness 10000\relax}{}{}

\graphicspath{ {Images/} }

\newcommand{\rd}{\, \mathrm{d}}

\newcommand{\be}{\begin{equation}\label}
\newcommand{\ee}{\end{equation}}
\newcommand{\bea}{\begin{eqnarray}\label}
\newcommand{\eea}{\end{eqnarray}}

\makeatletter
\newcommand*{\textoverline}[1]{$\overline{\hbox{#1}}\m@th$}
\newcommand*\bigcdot{\mathpalette\bigcdot@{.65}}
\newcommand*\bigcdot@[2]{\mathbin{\vcenter{\hbox{\scalebox{#2}{$\m@th#1\bullet$}}}}}
\makeatother

\newcommand{\eq}[1]{\begin{equation}#1\end{equation}}
\newcommand{\eqs}[1]{\begin{equation}\begin{split}#1\end{split}\end{equation}}
\date{}

\makeatletter

\allowdisplaybreaks
\numberwithin{equation}{section}
\author[1]{C. Armstrong\thanks{connor.armstrong@durham.ac.uk}}
\author[2]{H. Goodhew\thanks{hfg23@cam.ac.uk}}
\author[1]{A. Lipstein\thanks{arthur.lipstein@durham.ac.uk}}
\author[1]{J. Mei\thanks{jiajie.mei@durham.ac.uk}}

\affil[1]{Department of Mathematical Sciences, Durham University,
\authorcr  Stockton Road, DH1 3LE, Durham, United Kingdom}
\affil[2]{Department of Applied Mathematics and Theoretical Physics, University of Cambridge,
\authorcr  Wilberforce Road, Cambridge, CB3 0WA, UK}

\makeatother

\begin{document}

\title{Graviton Trispectrum from Gluons}

\maketitle

\begin{abstract}
The tree-level wavefunction coefficient for four gravitons in de Sitter space was recently bootstrapped using the Cosmological Optical Theorem, flat space limit, and Manifestly Local Test \cite{Bonifacio:2022vwa}. Inspired by the double copy for scattering amplitudes, we derive a compact new expression for this quantity starting from the wavefunction coefficient for gluons.  
\end{abstract}

\pagebreak

\tableofcontents

\section{Introduction}

Computing gravitational scattering amplitudes using standard Feynman diagram techniques is a formidable task due to the enormous number of terms that arise. On the other hand, modern approaches make use of a remarkable relation known as the double copy, which allows one to reduce gravitational calculations to much simpler calculations in gauge theory \cite{Zhu:1980sz,Kawai:1985xq,Bern:2008qj,Bjerrum-Bohr:2009ulz,Stieberger:2009hq,Bern:2010yg,Feng:2010my,Bern:2010ue, Carrasco:2011mn, Bern:2012uf}. Roughly speaking, it relates gravitational amplitudes to the square of gauge theory amplitudes. The double copy was first discovered in string theory, but applies to general theories of gravity coupled to matter, providing deep theoretical insights into the mathematical structure of gauge theory and gravity as well as powerful new computational tools which have important applications to the study of gravitational waves. For a review of recent developments, see \cite{Bern:2019prr,Adamo:2022dcm}.

By comparison, the study of perturbative gravitational observables analogous to scattering amplitudes is far less understood in curved backgrounds. Of particular interest are boundary correlators of gravitons in Anti-de Sitter space (AdS) and de Sitter space (dS), which play a prominent role in the AdS/CFT correspondence \cite{Maldacena:1997re} and cosmology \cite{Maldacena:2011nz,Strominger:2001gp,Maldacena:2002vr,McFadden:2009fg,McFadden:2010vh,Raju:2011mp}, respectively. In the context of cosmology, these quantities are known as wavefunction coefficients \cite{Hartle:1983ai} and cosmological correlators (or in-in correlators) can be obtained by squaring wavefunctions and computing expectation values \cite{Weinberg:2005vy,Maldacena:2002vr}. While there has been impressive progress in computing supergravity correlators in AdS using conformal bootstrap techniques \cite{Hartman:2022zik,Bissi:2022mrs}, it is not straightforward to adapt these methods to more realistic models in four dimensional de Sitter space (dS$_4$). Moreover, perturbative calculations in (A)dS encounter similar difficulties to those in flat space but are even more challenging due to the intrinsic complexity of working in curved backgrounds. Indeed, the tree-level wavefunction of four gravitons in dS$_4$ was only determined in full generality about four months ago \cite{Bonifacio:2022vwa} (see for \cite{Fu:2015vja} for earlier partial results). Despite the fact that Witten diagrams give hundreds of thousands of terms, the final result was only about a page in length. This simplification was achieved by using a powerful set of constraints including the flat space limit \cite{Maldacena:2011nz,Raju:2012zr}, Cosmological Optical Theorem (COT) \cite{Goodhew:2020hob,Melville:2021lst} and Manifestly Local Test (MLT) \cite{Jazayeri:2021fvk}, which are part of a broader arsenal of techniques collectively known as the cosmological boostrap \cite{Baumann:2022jpr}. Recent developments in this direction include geometric approaches \cite{Arkani-Hamed:2017fdk,Bzowski:2020kfw}, and methods based on factorisation \cite{Arkani-Hamed:2015bza,Arkani-Hamed:2018kmz,Baumann:2020dch,Baumann:2021fxj}, unitarity \cite{Meltzer:2021zin,Meltzer:2020qbr,Goodhew:2020hob,Melville:2021lst,Jazayeri:2021fvk,Goodhew:2021oqg}, Mellin-Barnes representations \cite{Sleight:2019hfp,Sleight:2021plv}, recursion relations \cite{Raju:2012zs,Armstrong:2022mfr,Albayrak:2023jzl}, color/kinematics duality \cite{Armstrong:2020woi,Albayrak:2020fyp,Alday:2021odx,Diwakar:2021juk,Sivaramakrishnan:2021srm,Cheung:2022pdk,Herderschee:2022ntr,Drummond:2022dxd}, scattering equations \cite{Roehrig:2020kck,Eberhardt:2020ewh,Gomez:2021qfd,Gomez:2021ujt}, and the double copy \cite{Farrow:2018yni,Li:2018wkt,Lipstein:2019mpu,Jain:2021qcl,Zhou:2021gnu,Armstrong:2022csc,Lee:2022fgr,Bissi:2022wuh}.

In this paper, we will combine the boostrap techniques used in \cite{Bonifacio:2022vwa} with the double copy, leading to a further reduction of the 4-graviton wavefunction down to only a few lines. Starting with the tree-level wavefunction for four gluons, we will apply a squaring procedure inspired by the double copy for flat space amplitudes. The resulting formula for the s-channel contribution to the wavefunction in \eqref{gravitonansatz} can be written in two lines and satisfies the flat space limit, COT, and MLT. Moreover it captures the vast majority of the hundreds of thousands of terms that arise from Witten diagrams. The full result for the s-channel contribution to the graviton wavefunction in \eqref{psigamma} can then be obtained by noting that the double copy ansatz contains spurious poles which can be cancelled by adding a simple two-line correction whose structure is fixed by the MLT. Morever, this correction can be deduced by looking at scalars exchanging a graviton. Using the double copy as a starting point, no new corrections arise after generalising this example to the gravitational case. Hence, while we do not yet have a systematic understanding of the double copy in (A)dS, it appears to be a very useful tool in the study of gravitational correlators.

This paper is organised as follows. In section \ref{review} we review some basics about the double copy and cosmological correlators. In section \ref{scalarwav} we derive the wavefunction for scalars exchanging a graviton starting from scalars exchanging a gluon. In section \ref{spinningwav} we generalise this procedure to derive a compact new formula for the wavefunction of four gravitons. In section \ref{conclusion} we discuss our conclusions and future directions. We also provide two Appendices summarising our notation and providing details on conformal time integrals.

\section{Review} \label{review}

In this section, we will review some facts about scattering amplitudes and cosmological correlators which will be useful later on.

\subsection{Double Copy}

Let us first review the double copy for scattering amplitudes in flat space. For simplicity, we will focus on tree-level 4-point amplitudes. A 4-point color-dressed gluon amplitude can be written as
\begin{align}
    \mathcal{A}_4=\frac{n_s c_s}{s} +\frac{n_t c_t}{t} +\frac{n_u c_u}{u},
\label{colordressed}
\end{align}
where we have set the gluon coupling to one, $s,t,u$ are Mandelstam variables:
\eqs{
s &=\left(k_{1}+k_{2}\right)^{\mu}\left(k_{1}+k_{2}\right)_{\mu}=2k_{1}^{\mu}k_{2\mu},\\
t &=\left(k_{1}+k_{4}\right)^{\mu}\left(k_{1}+k_{4}\right)_{\mu}=2k_{1}^{\mu}k_{4\mu},\\
u &=\left(k_{1}+k_{3}\right)^{\mu}\left(k_{1}+k_{3}\right)_{\mu}=2k_{1}^{\mu}k_{3\mu},
}
$n_{i}$ are kinematic numerators, and $c_i$ are color factors obeying the Jacobi relation:
\begin{equation}
c_{s}+c_{t}+c_{u}=0.
\label{jacobic}
\end{equation}
If we express $c_t$ in terms of $c_s$ and $c_u$ using \eqref{jacobic}, then \eqref{colordressed} can be written as
\begin{equation}
\mathcal{A}_{4}=c_{s} A_{1234}-c_{u} A_{1342},
\end{equation}
where the color-ordered amplitudes are given by
\eqs{
A_{1234}&=\frac{n_{s}}{s}-\frac{n_{t}}{t},\\
A_{1324}&=\frac{n_{t}}{t}-\frac{n_{u}}{u}.
\label{eq:ymamp}
}

The numerators are related by exchanges:
\eq{
n_t = -n_s\big|_{2\leftrightarrow4},\qquad n_u = -n_s\big|_{2\leftrightarrow3},
\label{eqn:numeratorexchange}
}
and obey an analogue of the Jacobi relation in \eqref{jacobic}:
\begin{equation}
n_{s}+n_{t}+n_{u}=0,\label{eq:kinjacobi}
\end{equation}
which is known as the kinematic Jacobi relation and encodes color/kinematics duality \cite{Bern:2008qj}. The double copy states that gravitational amplitudes can be obtained from color-dressed gluon amplitudes by replacing the color factors with kinematic numerators:
\begin{equation}
\mathcal{M}_{4}=\frac{n_{s}^{2}}{s}+\frac{n_{t}^{2}}{t}+\frac{n_{u}^{2}}{u},
\label{gravdouble}
\end{equation}
where we have set the gravitational coupling to 1. The double copy has been shown to hold for any multiplicity at tree-level \cite{Bern:2010yg,Feng:2010my} and to a very high order at loop level \cite{Bern:2010ue,Carrasco:2011mn,Bern:2012uf}. 

Generalised dimensional reduction \cite{Cachazo:2014xea} of the above gluon and graviton amplitudes implies a double copy for scalars exchanging gluons and gravitons, respectively. The basic idea is that $d$-dimensional scalars arise from $(d+1)$-dimensional polarisation vectors which point along the internal direction and are therefore orthogonal to $d$-dimensional momenta. In particular, writing the gravity polarisations in terms of polarisation vectors and taking the polarisation vectors to satisfy $\epsilon_a^\mu \epsilon_{b,\mu} = 1$ and $k_a^\mu \epsilon_{b,\mu}=0$ (where $a \neq b$ are particle labels), the first line of \eqref{eq:ymamp} reduces to
\eq{
\mathcal{A}^{1234}_{\phi} = \frac{t-u}{s} -\frac{u-s}{t},
}
which describes massless adjoint scalars exchanging a gluon. From this expression and \eqref{eqn:numeratorexchange} we can then read off that $n_s=t-u$, $n_t=u-s$, and $n_u=s-t$. Squaring the numerators according to \eqref{gravdouble} and noting that $s+t+u=0$ then gives
\eq{
\mathcal{M}^{\phi}_{4} = -4 \left(\frac{tu}{s} + \frac{us}{t} + \frac{st}{u}\right), 
}
which describes massless scalars exchanging a graviton and agrees with the generalised dimensional reduction of \eqref{gravdouble}. Note that the scalar amplitudes live in the same spacetime dimension as the gluon and graviton amplitudes, which is why we refer to this as generalised dimensional reduction.

\subsection{dS correlators}

Let us now switch our attention to cosmological correlators. We will work in the Poincar\'e patch of dS$_4$ with unit radius: 
\begin{equation}
{\rm d}s^{2}=(1/\eta)^{2}(-{\rm d}\eta^{2}+{\rm d}\vec{x}^{2}),
\label{metric}
\end{equation}
where $-\infty<\eta<0$ is the conformal time and $\vec{x}$ denotes the Euclidean boundary directions, with individual components $x^{i}$, $i=1,2,3$. Cosmological correlators (or in-in correlators) can be computed as follows:
\begin{equation}
\left\langle \phi(\vec{k}_{1})...\phi(\vec{k}_{n})\right\rangle =\frac{\int\mathcal{D}\phi \, \phi(\vec{k}_{1})...\phi(\vec{k}_{n})\left|\Psi\left[\phi\right]\right|^{2}}{\int\mathcal{D}\phi\left|\Psi\left[\phi\right]\right|^{2}},
\end{equation}
where $\phi$ represents the value of a generic bulk field in the future boundary Fourier transformed to momentum space, $\vec{k}_a$ are boundary momenta, and $\Psi\left[\phi\right]$ is the cosmological wavefuntion, which is a functional of $\phi$. For simplicity, we are considering a scalar field but in general, we should integrate over the boundary values of all the bulk fields, including the metric. 

The wavefunction can be perturbatively expanded as follows:
\begin{equation}
\ln\Psi\left[\phi\right]=-\sum_{n=2}^{\infty}\frac{1}{n!}\int\prod_{i=1}^{n}\frac{{\rm d}^{d}k_{i}}{(2\pi)^{d}}\psi_{n}\left(\vec{k}_{1},...\vec{k}_{n}\right)\phi(\vec{k}_{1})...\phi(\vec{k}_{n}),
\end{equation}
where the wavefunction coefficients $\psi_n$ can be expressed as
\begin{equation}
\psi_{n}=\delta^{d}(\vec{k}_T)\left\langle \left\langle \mathcal{O}\left(\vec{p}_{1}\right)...\mathcal{O}\left(\vec{p}_{n}\right)\right\rangle \right\rangle ,
\label{psid}
\end{equation}
where $\vec{k}_T=\vec{k}_{1}+...+\vec{k}_{n}$ and the object in double brackets can be treated as a CFT correlator in the future boundary \cite{Maldacena:2011nz,Bzowski:2012ih,Bzowski:2013sza,Arkani-Hamed:2015bza,Bzowski:2017poo,Arkani-Hamed:2018kmz,Bzowski:2018fql,Baumann:2020dch}. Note that momentum is conserved along the boundary but the total energy defined as 
\begin{equation}
E=\sum_{a=1}^n k_a, 
\end{equation}
where $k_a = |\vec{k}_a|$, is not conserved. The wavefunction coefficients in \eqref{psid} can be computed by analytically continuing AdS Witten diagrams and will be our main focus in this paper. In practice, we will drop the momentum conserving delta function when referring to the wavefunction coefficients. We will also analytically continue to Euclidean AdS when performing conformal time integrals.

For spinning fields we define the wavefunction coefficients in the helicity basis,
\begin{equation}
\ln\Psi\left[\gamma\right]=-\sum_{n=2}^{\infty}\frac{1}{n!}\int\prod_{i=1}^{n}\frac{{\rm d}^{d}k_{i}}{(2\pi)^{d}}\psi_{n}^{h_1\dots h_n}\left(\vec{k}_{1},...\vec{k}_{n}\right)\gamma^{h_1}(\vec{k}_{1})...\gamma^{h_n}(\vec{k}_{n}),
\end{equation}
where $h_a$ are helicities and are summed over. In order to apply the bootstrap methods outlined later in this section it is necessary to additionally define the so called ``trimmed'' wavefunction coefficients\cite{BBBB},
\begin{equation}\label{eq:trim}
\psi_{n}^{h_1\dots h_n}(\vec{k}_1\dots\vec{k}_n)=\sum_{\text{contractions}} \left[\epsilon_{1}^{h_{1}}\dots\epsilon_{n}^{h_{n}}\left(\vec{k}_{1}\right)^{\alpha_{1}}\dots\left(\vec{k}_{n}\right)^{\alpha_{n}}\right]\tilde{\psi}_n(\vec{k}_1,\dots,\vec{k}_n),
\end{equation}
where $\left(\vec{k}_{a}\right)^{\alpha_a}$ denotes the tensor product of $\alpha_a$ copies of $\vec{k}_a$, whose indices contract with those of the polarisation tensors on the left. The sum tells us that generically each wavefunction coefficient will contain several such trimmed terms and each one of these must be determined individually in the bootstrap approach. In the next two subsections we will describe how to compute wavefunction coefficients using Witten diagrams and boostrap techniques. 

\subsubsection{Witten Diagrams}
\label{sec:WD}
Our goal in this paper will be to deduce gravitational wavefunction coefficients from gluonic ones, so let us describe how to compute the latter in more detail. We will use Feynman rules in axial gauge in AdS momentum space first derived in \cite{Liu:1998ty,Raju:2011mp}. For notational simplicity, we will adopt conventions where factors of $i$ will not appear in the Feynman rules. See \cite{Albayrak:2018tam,Albayrak:2019asr,Albayrak:2019yve} for recent four- and five-point calculations using these Feynman rules. In axial gauge, gluons have the following bulk-to-boundary and bulk-to-bulk propagators in momentum space:
\begin{equation}
    G^A_{ij} (z,z',\vec{k})=-\int_{0}^{\infty}\omega d\omega\frac{z^{\frac{1}{2}}J_{\frac{1}{2}}(\omega z)J_{\frac{1}{2}}(\omega z')(z')^{\frac{1}{2}}}{k^{2}+\omega^{2}}H_{ij},
\end{equation}
where $\vec{k}$ is the momentum flowing through the propagator along the boundary directions, $k=|\vec{k}|$, $J_{\nu}$ is a Bessel function of the first kind, and
\eq{
H_{ij}=\eta_{ij}+\frac{k_{i}k_{j}}{\omega^{2}},
}
where $\eta_{ij}$ is the Euclidean boundary metric. Note that we have Wick rotated $\eta\rightarrow i z$, where $0 < z < \infty$, in order to make conformal time integrals manifestly convergent. The bulk-to-boundary propagator is given by
\begin{equation}
    G^A_i(z,\vec{k})=\epsilon _i \sqrt{\frac{2 k}{\pi}} z^{\frac{1}{2}}{K} _{\frac{1}{2}}(kz),
\end{equation}
where $\vec{k}$ and $\vec{\epsilon}$ are the boundary momentum and polarisation vector, respectively, which satisfy $\epsilon\cdot \epsilon=\epsilon\cdot k=0$ (where the dot denotes an inner product of 3-vectors), and $K_{\nu}$ is a modified Bessel function of the second kind. 

The color-ordered Feynman vertices for gluons have the same structure as in flat space but the indices only run over the boundary directions in axial gauge. In more detail, the three and four-point vertices are
\eqs{
    V_{jkl}(\vec{k}_1,\vec{k}_2,\vec{k}_3) &= \left(\eta _{jk}(\vec{k}_1-\vec{k}_2)_l +\eta _{kl}(\vec{k}_2-\vec{k}_3)_j+\eta _{lj}(\vec{k}_3-\vec{k}_1)_k\right),\\
    V_{jklm}&=2\eta_{jl}\eta_{km}-\left(\eta_{jk}\eta_{lm}+\eta_{jm}\eta_{kl}\right),
\label{ymvertices}
}
where we have set the gluon coupling $g=\sqrt{2}$ for convenience. When computing color-ordered 4-point wavefunctions, it will be convenient to split the 4-point contact diagram into an s and t-channel contribution. After dressing the second line of \eqref{ymvertices} with polarisations we then get the following quantities:
\eqs{
V_{c}^{s} &=\epsilon_{1}\cdot\epsilon_{3}\epsilon_{2}\cdot\epsilon_{4}-\epsilon_{1}\cdot\epsilon_{4}\epsilon_{2}\cdot\epsilon_{3},\\
V_{c}^{t} &=\epsilon_{1}\cdot\epsilon_{2}\epsilon_{3}\cdot\epsilon_{4}-\epsilon_{1}\cdot\epsilon_{3}\epsilon_{2}\cdot\epsilon_{4}.
}
Finally, we note that for each interaction vertex, we must perform an integral over the AdS radius along with the measure $\sqrt{\det{g}}=z^{-4}$. In practice there will be additional factors of $z$ coming from the inverse metrics used to contract indices. 

Although we will not need them in this paper, the bulk-to-boundary and bulk-to-bulk propagators for gravitons in axial gauge are given by
\begin{align}
G_{i j}^{\gamma}(z,\vec{k})=& \epsilon_{i j} \sqrt{\frac{2}{\pi}} z^{-2}(k z)^{\frac{3}{2}} K_{\frac{3}{2}}(k z), \\
G_{ij,kl}^{\gamma}\left(z, z^{\prime},\vec{k}\right)=& \frac{-\left(z z^{\prime}\right)^{-\frac{1}{2}}}{2} \int_{0}^{\infty} d \omega J_{\frac{3}{2}}(\omega z) J_{\frac{3}{2}}\left(\omega z^{\prime}\right) \frac{\omega\left(H_{ik}H_{jl}+H_{il}H_{jk}-H_{ij}H_{kl}\right)}{k^{2}+\omega^{2}},
\label{gravitonprop}
\end{align}
where $\epsilon_{ij}=\epsilon_i \epsilon_j$ is a graviton polarisation. The Feynman rules for scalars coupled to gluons and gravitons can then be deduced by setting $\epsilon_a \cdot \epsilon_b=1$ and $\epsilon_a \cdot k_b=0$, where $a\neq b$ and the polarisations correspond to external scalars. For example, the scalar bulk-to-boundary propagator is 
\eq{
G^{\phi} (z,\vec{k}) = \sqrt{\frac{2}{\pi}} z^{3 / 2} k^{\nu} K_{\nu}(k z),
}
where $\nu=1/2$ for conformally coupled scalars (which descend from gluons) and $\nu=3/2$ for massless scalars (which descend from gravitons). Moreover, the three-point scalar-scalar-gluon vertex can be deduced from the first line of \eqref{ymvertices} by dressing two of the legs with polarisations and performing the generalised dimensional reduction procedure described above:
\eq{
v_i(\vec{k}_1,\vec{k}_2,\vec{k}_3) = (k_1-k_2)_i,
}
where leg three is a gluon with index $i$.

\subsubsection{Bootstrap}
While it is relatively straightforward to compute 4-point gluon wavefunctions using Witten diagrams, doing so for gravitons is very challenging due to the large number of terms that arise. It was shown in \cite{Bonifacio:2022vwa} that the graviton trispectrum is completely fixed up to arbitrary (non-local) field redefinitions by the combination of the flat space limit\cite{Maldacena:2011nz,Raju:2012zr}, the Cosmological Optical Theorem (COT)\cite{Goodhew:2020hob,Melville:2021lst} and the Manifestly Local Test (MLT)\cite{Jazayeri:2021fvk}. These tools have been established over the last few years as key ingredients in the Cosmological Bootstrap\cite{Baumann:2022jpr}. A consequence of this is that any expression that satisfies the COT and has the correct flat space limit can be combined with the MLT to give the graviton trispectrum. To aid the reader we will briefly review these three tests.

Fields in the Bunch-Davies vacuum in the infinite past of de Sitter behave just like flat space fields. As a result, wavefunction coefficients contain the flat space amplitude within them as the residue of the total energy pole. For Einstein gravity (being a two derivative theory) this means that
\begin{align}
    \lim_{E\rightarrow 0} \psi^{\gamma}_4 \propto \frac{k_1k_2k_3k_4}{E^3}\mathcal{M}_4,
\end{align}
where $E=k_1+k_2+k_3+k_4$ and $\mathcal{M}_4$ is the 4-graviton amplitude. While a naive squaring of the tree-level 4-point gluonic wavefunction satisfies the correct flat space limit \cite{Armstrong:2020woi}, it does not satisfy the COT, which we describe in the next paragraph. To remedy this, we will instead consider squaring the numerators in the conformal time integrand. 

As a consequence of unitary time evolution in the bulk de Sitter space time, all wavefunction coefficients satisfy the so-called COT. This relationship relates exchange diagrams to simpler diagrams involving one fewer exchanged particle. In the case of gravity, this relationship can be expressed as
\begin{align}\label{cot}
    {\psi}&_4^{h_1h_2h_3h_4}(k_1,k_2,k_3,k_4,k_s,k_t)+\psi_4^{h_1h_2h_3h_4}(-k_1,-k_2,-k_3,-k_4,k_s,-k_t)^*= \nonumber \\&\sum_{h}P^h(k_s)\left[\psi_3^{h_1h_2h}(k_1,k_2,k_s)-\psi_3^{h_1h_2h}(k_1,k_2,-k_s)\right]\left[\psi^{h_3h_4h}_3(k_3,k_4,k_s)-\psi^{h_3h_4h}_3(k_3,k_4,-k_s)\right],
\end{align}
where $k_s=|\vec{k}_{1}+\vec{k_{2}}|$ and $k_t=|\vec{k}_{1}+\vec{k}_{4}|$. Note that the $k_t$ dependence in the left-hand-side is encoded by the polarisation sum on the right-hand-side. We also note that there will be some dependence on the directions of the momenta through the polarisation tensors but this has been left implicit due to the convention that they are unchanged when we adjust the energies \cite{Goodhew:2021oqg}. As was noticed in \cite{Jazayeri:2021fvk} this is sufficient to fix all of the partial energy poles and so any result satisfying both this and the flat space limit will be equal to the full answer up to sub-leading total energy poles.

Finally, any four-point\footnote{Equivalent results exist for more general interactions but the expression given here is explicitly for a 4-point wavefunction coefficient.} interaction arising from a Lagrangian with no inverse Laplacian acting on single fields (such as that arising from Einstein gravity) must generate a wavefunction coefficient that satisfies the so-called MLT:
\begin{equation}\label{mlt}
    \lim_{k_1\rightarrow 0}\partial_{k_1}\tilde{\psi}_4(k_1,k_2,k_3,k_4,k_s,k_t)=0,
\end{equation}
which is true even away from physical momentum configurations (unlike, for example, the soft theorems). The tilde indicates that this applies to the trimmed wavefunction coefficients as the kinematics of the polarisation tensors can introduce poles in the wavefunction that violate the assumptions that go into the MLT. As we will see later, our double copy prescription will satisfy the flat space limit, COT, and MLT, but will contain spurious poles requiring us to add a simple correction whose structure will be fixed by the MLT.

\section{Scalar Wavefunctions} \label{scalarwav}

In this section, we will derive a compact new formula for the 4-point wavefunction of minimally coupled scalars exchanging a graviton starting from the wavefunction for conformally coupled scalars exchanging a gluon. This will be a warm-up for obtaining the 4-point graviton wavefunction from the gluonic one in the next section. Indeed, the scalar wavefunctions we derive in this section can be obtained via generalised dimensional reduction of the spinning ones.

\subsection{Ansatz}
\begin{figure}
    \centering
    \includegraphics{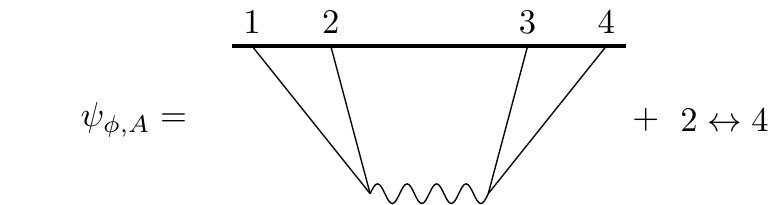}
    \caption{Witten diagrams for conformally coupled scalars exchanging a gluon.}
    \label{fig:ConfScalar}
\end{figure}
Let us begin with conformally coupled scalars exchanging a gluon. We will consider the color-ordered wavefunction analogous to the first line of \eqref{eq:ymamp} in flat space and take the scalars to be in the adjoint representation of the gauge group. Using the Feynman rules in \ref{sec:WD}, the $s$-channel Witten diagram depicted in Figure \ref{fig:ConfScalar} is given by
\eq{
\psi_{\phi,A}^{(s)}= \int\frac{d\omega\,\omega}{k_s^2+\omega^2}dz\,dz'\,(KKJ)^{1/2}_{12}(z)(KKJ)^{1/2}_{34}(z')N_s^{\phi},
}
where the integrals over $\omega$, $z$, and $z'$ are from zero to infinity, $k_s=|\vec{k}_1+\vec{k}_2|$,
\begin{equation}
N_s^{\phi}=v_{12}^i v_{34}^j H_{ij},
\label{phinum}
\end{equation}
$v_{12}^i=(\vec{k}_1-\vec{k}_2)^i$, $H_{ij}=\eta_{ij}+\frac{\vec{k}_{12}^i\vec{k}_{12}^j}{\omega^2}$, and $\vec{k}_{ab}=\vec{k}_{a}+\vec{k}_{b}$. The $KKJ$ integrals are given by
\eqs{
(KKJ)^{\nu}_{ab} = \frac{2}{\pi}(k_ak_bz)^\nu zK_{\nu}(k_az)K_{\nu}(k_bz)J_{\nu}(\omega z).
}
We have also dropped the overall factor of $i$. 

The numerator $N_s^{\phi}$ can be thought of as the analogue of the kinematic numerator $n_s$ in \eqref{eq:ymamp}. By analogy to \eqref{gravdouble} a natural guess for the double copy is
\begin{equation}
\psi_{\phi,\mathrm{DC}}^{(s)}\overset{?}{=}\int\frac{d\omega\,\omega}{k_{s}^{2}+\omega^{2}}d z\,dz'\,(KKJ)_{12}^{3/2}(z)(KKJ)_{34}^{3/2}(z')\left(N_{s}^{\phi}\right){}^{2},
\end{equation}
where we have replaced $\nu=\frac{1}{2}$ Bessel functions with $\nu=\frac{3}{2}$ Bessel functions, as expected for mininimally coupled scalars and gravitons, and squared the numerator. While this guess has the correct flat space limit, it does not satisfy the COT in \eqref{cot}. Looking at the graviton bulk-to-bulk propagator in \eqref{gravitonprop} then motivates the following ansatz:
\eqs{
\psi_{\phi,\mathrm{DC}}^{(s)}=\int\frac{d\omega\,\omega}{k_{s}^{2}+\omega^{2}}dz\,dz'\,(KKJ)_{12}^{3/2}(z)(KKJ)_{34}^{3/2}(z')\left(\left(N_{s}^{\phi}\right){}^{2}-\frac{1}{2}\tilde{v}_{12}^{ij}H_{ij}\tilde{v}_{34}^{kl}H_{kl}\right),
\label{ansatzscalar}
}
where $\tilde{v}_{12}^{ij}=v_{12}^iv_{12}^j$. While the second term in parenthesis is similar in structure to the third term in \eqref{gravitonprop}, it is constructed from scalar-scalar-gluon vertices. We will say more about the double copy origin of this term in section \ref{doublecopystruct}. In Appendix \ref{integrals}, we evaluate the integrals in \eqref{ansatzscalar} and obtain a more explicit formula:
\eqs{
\psi_{\phi,\mathrm{DC}}^{(s)} &= \frac{1}{3}k_s^4f_{2,2}\Pi_{2,2} - \frac{1}{3}k_s^2k_{12}k_{34}f_{2,1}\Pi_{2,1} + \frac{1}{2}f_{2,0}\frac{k_{12}^2\alpha^2k_{34}^2\beta^2}{k_s^4}\\
&\qquad-\frac{1}{2}f_{2,1}\left(\left(k_{12}^2+\alpha^2-k_s^2-\frac{k_{12}^2\alpha^2}{k_s^2}\right)\frac{k_{34}^2\beta^2}{k_s^2} + \frac{k_{12}^2\alpha^2}{k_s^2}\left(k_{34}^2+\beta^2-k_s^2-\frac{k_{34}^2\beta^2}{k_s^2}\right)\right),
\label{psiphidc}
}
where $k_{ij}=k_i+k_j$, $\alpha=k_1-k_2$, $\beta=k_3-k_4$, $\Pi_{2,2}$ and $\Pi_{2,1}$ are polarisation sums given in \eqref{psisum}, and $f_{2,2}$, $f_{2,1}$, and $f_{2,0}$ are conformal time integrals given in \eqref{integrals1}.

\subsection{Corrections}

While the ansatz in \eqref{ansatzscalar} has the correct flat space limit and satisfies the COT, after integration we find that it contains spurious poles in $k_{12}$ and $k_{34}$. These can be cancelled by adding the following simple correction:
\eq{\label{eq:SP}
\psi_{\rm{sp}}^{(s)}= -\frac{1}{2}\left(\frac{2 k_1 k_2 k_3 k_4}{\left(k_{12}+k_{34}\right)^2}\left(\frac{\alpha^2}{k_{34}}+\frac{\beta ^2}{k_{12}}\right)+\frac{\alpha ^2 k_3 k_4}{k_{34}}+\frac{\beta ^2 k_1 k_2}{k_{12}}\right).
}
In fact, this is the unique correction which cancels the spurious poles without affecting the flat space limit or COT, modulo adding terms which do not contain spurious poles. This ambiguity can be fixed by the MLT, which is satisfied by \eqref{ansatzscalar} but not \eqref{eq:SP}.

Following the procedure in \cite{Bonifacio:2022vwa}, we will construct an ansatz for the missing terms and fix it by enforcing the MLT. As was shown in \cite{Anninos:2014lwa,Goodhew:2022ayb}, the most general tree-level wavefunction coefficient for interactions involving Einstein gravity is a rational function of the energies. Moreover, the correction terms can have at worst $E^{-2}$ poles and no other singularities so as not to affect the flat space limit or COT \footnote{We are free to add in some non-local field redefinitions with $k_s$ poles, as was shown in\cite{Bonifacio:2022vwa}, but these are always present and so will be ignored in our ansatz.}. Scale invariance also forces any correction term to scale like momentum cubed so the most general correction must have the form 
\begin{equation}
   \psi_{\rm{MLT}}^{(s)}=\frac{\textrm{Poly}^{(5)}(k_1,k_2,k_3,k_4,k_s,k_t,k_u)}{E^2},
\end{equation}
where $\textrm{Poly}^{(5)}$ is a general polynomial with homogeneity degree $5$ under rescaling momenta. We can simplify this by noting that we are only adjusting the s-channel and so anything that we add must respect the $s$-channel symmetries. To encode the $k_1\leftrightarrow k_2$ and $k_3\leftrightarrow k_4$ exchange symmetry we express this polynomial as a function of the combinations $k_{12},\,k_1 k_2,\,k_{34},\,k_3 k_4$ and $k_s^2$. The remaining dependence on $k_t$ and $k_u$ can only be through the combination $k_t^2-k_u^2$, which picks up a minus sign under $k_1\leftrightarrow k_2$ and so must be multiplied by something else that also behaves in this way. Therefore, 
\begin{align}
    \psi_{\rm{MLT}}^{(s)}=\frac{\textrm{Poly}^{(5)}(k_{12},k_1k_2,k_{34},k_3k_4,k_s^2)+A_1 \alpha\beta(k_t^2-k_u^2)E+A_2(k_t^2-k_u^2)^2 E}{E^2}+\left(\vec{k}_{1},\vec{k}_{2}\right)\leftrightarrow\left(\vec{k}_{3},\vec{k}_{4}\right),
\label{mltcorr}
\end{align}
where the contribution at the end is required to recover the $s$-channel symmetry that is not explicit in the construction of this polynomial. This ansatz has a total of 17 free coefficients.
 
On fixing the free coefficients such that \eqref{mltcorr} combines with \eqref{eq:SP} to satisfy the MLT we find 
\eqs{
     \psi_{\rm{MLT}}^{(s)}&=A(k_1^3+k_2^3+k_3^3+k_4^3)+\frac{1}{2E}\Big( (k_1k_3+k_2k_4)(k_1k_4+k_2k_3)-2(\alpha^2k_1k_2+\beta^2k_3k_4)\\&\qquad+(\alpha^2k_3k_4+\beta^2k_1k_2)
    -3(k_{34}^2k_1k_2+k_{12}^2k_3k_4)+2(k_{12}^2k_1k_2+k_{34}^2k_3k_4)+6k_1k_2k_3k_4\\
    &\qquad+k_{12}k_{34}(E^2-2(k_1k_2+k_3k_4))\Big),
}
where $A$ is a free coefficient that corresponds to the field redefinition $\phi\rightarrow \phi+A\phi^3$, where $\phi$ is the external scalar field. Choosing $A=-7/2$ then gives the compact form
\eqs{
\psi_{\rm{MLT}}^{(s)} &= \frac{5k_1k_2k_3k_4}{E}+\frac{E}{2}(k_{12}k_{34}-4k_1k_2-4k_3k_4)-\frac{1}{E}(k_1k_2-k_3k_4)(\alpha^2-\beta^2)-3(\alpha^2k_{12}+\beta^2k_{34}).
\label{psimlt}
}

In summary, we find that the s-channel contribution to the wavefunction for minimally coupled scalars exchanging a graviton can be written as
\begin{equation}
\psi_{\phi,\gamma}^{(s)}=\psi_{\phi,\mathrm{DC}}^{(s)}+\psi_{\rm{sp}}^{(s)}+\psi_{\rm{MLT}}^{(s)},
\label{psiphigamma}
\end{equation}
where the three terms on the right-hand-side are given by \eqref{ansatzscalar}, \eqref{eq:SP}, and \eqref{psimlt}. The full wavefunction can then be obtained by summing over all three channels, where the contributions from the $t$ and $u$ channels can be obtained from \eqref{psiphigamma} by exchanging $2\leftrightarrow3$ and $2\leftrightarrow4$. More explicitly, plugging in \eqref{psiphidc} we obtain
\eqs{
\psi_{\phi,\gamma}^{(s)} &=  \frac{1}{3}k_s^4f_{2,2}\Pi_{2,2} - \frac{1}{3}k_s^2k_{12}k_{34}f_{2,1}\Pi_{2,1} + \frac{1}{2}k_{12}k_{34}(k_{12}k_{34}+k_s^2)f_{2,1}\left(-\frac{\alpha^2+\beta^2}{k_s^2}+3\frac{\alpha^2}{k_s^2}\frac{\beta^2}{k_s^2}\right)\\&\qquad - \frac{1}{2}\frac{k_s^2}{E}(k_1k_2+k_3k_4+E^2)\frac{\alpha^2}{k_s^2}\frac{\beta^2}{k_s^2}-\frac{1}{2E}(k_1k_2-k_3k_4)(\alpha^2-\beta^2)-\frac{5}{2}(k_{12}\alpha^2+k_{34}\beta^2)\\
&\qquad +\frac{5k_1k_2k_3k_4}{E}+\frac{E}{2}\left(k_{12}k_{34}-4k_1k_2-4k_3k_4\right).
\label{scalarfull}
}
This agrees up to field redefinition with the result previously obtained in \cite{Bonifacio:2022vwa} using bootstrap methods.

\section{Spinning Wavefunctions} \label{spinningwav}

In this section, we will generalise the procedure in the previous section to deduce the tree-level 4-point graviton wavefunction from gluons, arriving at a compact new formula.

\subsection{Ansatz and Corrections}
\begin{figure}
    \centering
    \includegraphics{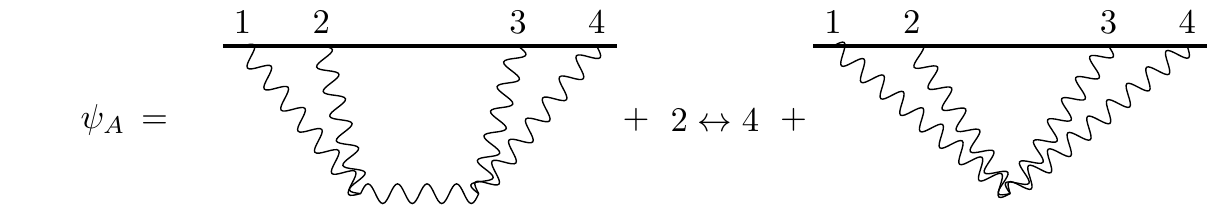}
    \caption{Witten diagrams for four-point gluon wavefunction coefficient.}
    \label{fig:Gluon}
\end{figure}
Let us start with the s-channel contribution to the 4-point color-ordered gluon wavefunction, depicted in Figure \ref{fig:Gluon}. Using the Feynman rules in section \ref{sec:WD} we obtain
\eqs{
\psi_{A}^{(s)}&=\int\frac{dz dz'd\omega\omega}{\left(k_{s}^{2}+\omega^{2}\right)}\left(KKJ\right)_{12}^{1/2}(z)\left(KKJ\right)_{34}^{1/2}(z')V_{12}^{i}H_{ij}V_{34}^{j}\\&\qquad+\int dz dz'\delta(z-z')\left(KK\right)_{12}^{1/2}(z)V_{c}^{s}\left(KK\right)_{34}^{1/2}(z')
\label{ymwitten}
}
where $\left(KK\right)_{ab}^{1/2}(z)=\sqrt{k_{a}k_{b}}z K_{1/2}\left(k_{a} z \right)K_{1/2}\left(k_{b} z\right)$,
\eqs{
V_{ab}^{i} &=\epsilon_{a}\cdot\epsilon_{b}\,(\vec{k}_{a}-\vec{k}_{b})^{i}+2\epsilon_{a}\cdot\vec{k}_{b}\epsilon_{b}^{i}-2\epsilon_{b}\cdot\vec{k}_{a}\epsilon_{a}^{i},\\
V_c^s &=\epsilon_1\cdot\epsilon_3\,\epsilon_2\cdot\epsilon_4-\epsilon_1\cdot\epsilon_4\,\epsilon_2\cdot\epsilon_3.
\label{eqn:GravDCRules}
}
The second term in \eqref{ymwitten} arises from a bulk contact interaction so we have written it as integral over two bulk points with a delta function. To combine it with the first term, use the orthogonality of Bessel functions
\begin{equation}
\delta(z-z')=\int d\omega\,\omega\left(zz'\right)^{1/2}J_{1/2}\left(\omega z\right)J_{1/2}\left(\omega z'\right).
\end{equation}
We then obtain
\eq{
\psi_{A}^{(s)}=\int\frac{d\omega\,\omega}{k_{s}^{2}+\omega^{2}}dz\,dz'\,(KKJ)_{12}^{1/2}(z)(KKJ)_{34}^{1/2}(z')N_{s},
}
where the numerator $N_s$ is
\eq{
N_{s}=V_{12}^{i}H_{ij}V_{34}^{j}+V_{c}^{s}(\omega^{2}+k_{s}^{2}).
}

By analogy with the scalar double copy ansatz in \eqref{ansatzscalar}, a natural guess for gravitons is
\eqs{
\psi_{\gamma,\mathrm{DC}}^{(s)}\overset{?}{=}\int\frac{d\omega\,\omega}{k_{s}^{2}+\omega^{2}}dz\,dz'\,(KKJ)_{12}^{3/2}(z)(KKJ)_{34}^{3/2}(z')\left(N_{s}^{2}-\frac{1}{2}\tilde{V}_{12}^{ij}H_{ij}\tilde{V}_{34}^{kl}H_{kl}\right),
}
where $\tilde{V}_{ab}^{ij}=V_{ab}^iV_{ab}^j$. While this ansatz satisfies the COT, the second term spoils the flat space limit. This can be remedied by adding one more term to the integrand:
\eqs{
\psi_{\gamma,\mathrm{DC}}^{(s)} &= \int\frac{d\omega\,\omega}{k_s^2+\omega^2}dz\,dz'\,(KKJ)^{3/2}_{12}(z)(KKJ)^{3/2}_{34}(z')\\&\qquad\times\left(N_s^2-\frac{1}{2}\tilde{V}^{ij}_{12}H_{ij}\tilde{V}_{34}^{kl}H_{kl} + \frac{1}{2}(\epsilon_1\cdot\epsilon_2)^2(\epsilon_3\cdot\epsilon_4)^2(\omega^2+k_s^2)^2\right).
\label{gravitonansatz}
}
In the next subsection, we will explain how the second and third terms arise from the double copy. 

As before, we must add terms to cancel spurious poles and satisfy the MLT. Remarkably, these turn out to be identical to the scalar case after dressing with polarisations. In the end, we find that the $s$-channel contribution to the 4-point graviton wavefunction is given by
\eqs{
\psi_{\gamma}^{(s)}=\psi_{\gamma,\mathrm{DC}}^{(s)}+\left(\epsilon_{1}\cdot\epsilon_{2}\right)^{2}\left(\epsilon_{3}\cdot\epsilon_{4}\right)^{2}\left(\psi_{\rm{sp}}^{(s)}+\psi_{\rm{MLT}}^{(s)}\right),
\label{psigamma}
}
where the terms on the right-hand-side are given in \eqref{gravitonansatz}, \eqref{eq:SP}, and \eqref{psimlt}. The full wavefunction can then be obtained by summing over all three channels, where the contributions from the $t$ and $u$ channels can be obtained from \eqref{psigamma} by exchanging $2\leftrightarrow3$ and $2\leftrightarrow4$. This non-trivially agrees with the result previously obtained in \cite{Bonifacio:2022vwa} using bootstrap methods, but now provides a more compact expression which exposes the underlying double copy structure. 

In Appendix \ref{integrals} we evaluate the integrals in \eqref{gravitonansatz} to obtain the following more explicit formula:
\eqs{
\psi_{\gamma}^{(s)} &= (\epsilon_1\cdot \epsilon_2)^2( \epsilon_3\cdot \epsilon_4)^2\psi^{(s)}_{\mathrm{DC}}+\left(8(\epsilon_1\cdot \epsilon_2)( \epsilon_3\cdot \epsilon_4)W_sk_s^2\Pi_{1,1}+16W_s^2\right)f_{2,2}\\&\qquad-(\epsilon_1\cdot \epsilon_2)( \epsilon_3\cdot \epsilon_4)k_{12}k_{34}\left(8W_s\Pi_{1,0}+\alpha\beta V^s_c\right)f_{2,1}+\left((V^s_c)^2+\frac{1}{2}(\epsilon_1\cdot \epsilon_2)^2( \epsilon_3\cdot \epsilon_4)^2\right)f_a\\
&\qquad+\left((\epsilon_1\cdot \epsilon_2 )(\epsilon_3\cdot \epsilon_4)(\vec{k}_1-\vec{k}_2)\cdot(\vec{k}_3-\vec{k}_4)+4W_s\right)V^s_cf_b,
\label{integratedpsigam}
}  
where $f_a$ and $f_b$ are given in \eqref{integrals2}, $V^s_c$ is given in \eqref{eqn:GravDCRules}, and the other tensor structure is given by
\eqs{
W_s &= (\epsilon_1\cdot\epsilon_2)\left[(k_1\cdot\epsilon_3)(k_2\cdot\epsilon_4)-(k_2\cdot\epsilon_3)(k_1\cdot\epsilon_4)\right]+(\epsilon_3\cdot\epsilon_4)\left[(k_3\cdot\epsilon_1)(k_4\cdot\epsilon_2)-(k_4\cdot\epsilon_1)(k_3\cdot\epsilon_2)\right]\\
&\qquad+\left[(k_2\cdot\epsilon_1)\epsilon_1-(k_1\cdot\epsilon_2)\epsilon_2\right]\cdot\left[(k_4\cdot\epsilon_3)\epsilon_4-(k_3\cdot\epsilon_4)\epsilon_3\right].
}
Notice that the tensor structure $(\epsilon_1\cdot\epsilon_3)(\epsilon_3\cdot \epsilon_2)(\epsilon_2\cdot \epsilon_4)(\epsilon_4\cdot \epsilon_1)$, which appeared in the result presented in \cite{Bonifacio:2022vwa}, is absent in \eqref{integratedpsigam} but this merely reflects a different choice of tensors to represent the answer. This contribution is instead captured by the $(V_c^s)^2$ and $V_c^s (\epsilon_1\cdot \epsilon_2)(\epsilon_3\cdot\epsilon_4)$ terms as well as a modification to the $(\epsilon_1\cdot \epsilon_2)^2(\epsilon_3\cdot\epsilon_4)^2$ term.

\subsection{Double Copy Structure} \label{doublecopystruct}

The simple ansatz in \eqref{gravitonansatz} captures most of the terms in the 4-point graviton wavefunction. Comparing this to \eqref{gravdouble}, we see that the analogue of the graviton numerator in the $s$-channel is
\begin{equation}
N_{s}^\gamma=N_{s}^{2}-\frac{1}{2}\tilde{V}_{12}^{ij}H_{ij}\tilde{V}_{34}^{kl}H_{kl}+\frac{1}{2}\left(\epsilon_{1}\cdot\epsilon_{2}\right)^2\left(\epsilon_{3}\cdot\epsilon_{4}\right)^{2}\left(k_{s}^{2}+\omega^{2}\right)^{2},\label{gravnum}
\end{equation}
where $N_s$ is the $s$-channel gluon numerator:
\begin{equation}
N_{s}=V_{12}^{\underline{i}}V_{34}^{j}H_{\underline{i}j}+\left(k_{s}^{2}+\omega^{2}\right)\epsilon_{12}^{\underline{ij}}\epsilon_{34}^{kl}\eta_{\underline{i}[k}\eta_{l]\underline{j}},\label{YMnum}
\end{equation} 
where $\epsilon_{12}^{\underline{ij}}=\epsilon_{1}^{\underline{i}}\epsilon_{2}^{\underline{j}}$ and we have added an underscore to indices associated with the left side of the s-channel Witten diagrams, i.e. legs 1 and 2.

Naively squaring the gluon numerator gives tensors which contract indices on the left with indices on the right:
\begin{equation}
N_{s}^{2}=\tilde{V}_{12}^{\underline{ij}} \tilde{V}_{34}^{kl}T_{\underline{ij}kl}+2\left(k_{s}^{2}+\omega^{2}\right)\epsilon_{12}^{\underline{ij}}V_{12}^{\underline{k}}\epsilon_{34}^{lm}V_{34}^{n}T_{\underline{ijk}lmn}+\left(k_{s}^{2}+\omega^{2}\right)^{2}\epsilon_{12}^{\underline{ij}}\epsilon_{12}^{\underline{kl}}\epsilon_{34}^{mn}\epsilon_{34}^{pq}T_{\underline{ijkl}mnpq},
\end{equation}
where
\eqs{
T_{\underline{ij}kl}&=H_{\underline{i}k}H_{\underline{j}l},\\
T_{\underline{ijk}lmn}&=\eta_{\underline{i}[l}\eta_{m]\underline{j}}H_{\underline{k}n},\\
T_{\underline{ijkl}mnpq}&=\eta_{\underline{i}[m}\eta_{n]\underline{j}}\eta_{\underline{k}[p}\eta_{q]\underline{l}}.
}
On the other hand, we can also consider an alternative prescription for squaring the tensor structures where indices on the left are never contracted with indices on the right:  
\eqs{
\tilde{N}_{s}^{2}\equiv\tilde{V}_{12}^{\underline{ij}} \tilde{V}_{34}^{kl}\tilde{T}_{\underline{ij}kl}+2\left(k_{s}^{2}+\omega^{2}\right)\epsilon_{12}^{\underline{ij}}V_{12}^{\underline{k}}\epsilon_{34}^{lm}V_{34}^{n}\tilde{T}_{\underline{ijk}lmn}+\left(k_{s}^{2}+\omega^{2}\right)^{2}\epsilon_{12}^{\underline{ij}}\epsilon_{12}^{\underline{kl}}\epsilon_{34}^{mn}\epsilon_{34}^{pq}\tilde{T}_{\underline{ijkl}mnpq},
\label{eq:numeratordouble}
}
where
\eqs{
\tilde{T}_{\underline{ij}kl}&=H_{\underline{ij}}H_{kl},\\
\tilde{T}_{\underline{ijk}lmn}&=0,\\
\tilde{T}_{\underline{ijkl}mnpq}&=\lambda\eta_{\underline{ij}}\eta_{\underline{kl}}\eta_{mn}\eta_{pq}.
\label{doublenew}
}
Note that $\lambda$ in the third line is an unfixed relative coefficient and the second line vanishes because there are an odd number of indices with or without underscores so there is no way to contract all of them. 

Hence we find that there are two ways to define the double copy of the gluon numerator. Moreover, consistency with the flat space limit and the COT implies that both are required. Indeed, \eqref{gravnum} can be written as 
\begin{equation}
N_{s}^{\gamma} = N_{s}^{2}-\frac{1}{2}\tilde{N}_{s}^{2},
\end{equation}
where we set $\lambda=-1$ in \eqref{doublenew}. This can be also written in terms of asymmetric products of deformed numerators:
\begin{equation}
N_{s}^{\gamma}= \frac{1}{2}\left(N_{12}^{-}N_{34}^{+}+N_{12}^{+}N_{34}^{-}\right),
\label{doublefinal}
\end{equation}
where
\eqs{
N_{12}^{\pm}&=N_{s}+\frac{i}{\sqrt{2}}\epsilon_{1}\cdot\epsilon_{2}\epsilon_{3}\cdot\epsilon_{4}\left(\omega^{2}+k_{s}^{2}\right)\pm \frac{1}{\sqrt{2}}\tilde{V}_{12}^{ij}H_{ij},\\
N_{34}^{\pm}&=N_{s}-\frac{i}{\sqrt{2}}\epsilon_{1}\cdot\epsilon_{2}\epsilon_{3}\cdot\epsilon_{4}\left(\omega^{2}+k_{s}^{2}\right)\pm \frac{1}{\sqrt{2}}\tilde{V}_{34}^{ij}H_{ij}.
\label{deformednum}
}
It would be interesting to explore if these numerators encode some analogue of color/kinematics duality.

This story can easily be adapted to the case of external scalars using generalised dimensional reduction, i.e. setting $\epsilon_a \cdot k_b=0$ and $\epsilon_a \cdot \epsilon_b=1$ for $a \neq b$. Indeed, dimensional reduction of the gluon numerator in \eqref{YMnum} gives
\begin{equation}
N_{s}^{\phi}=v_{12}^{\underline{i}}v_{34}^{j}H_{\underline{i}j}\label{scalarnum},
\end{equation}
which agrees with \eqref{phinum}. Applying the two double copy prescriptions described above then gives
\begin{equation}
\left(N_{s}^{\phi}\right)^{2}=\tilde{v}_{12}^{\underline{ij}}\tilde{v}_{34}^{kl}T_{\underline{ij}kl},\,\,\,\left(\tilde{N}_{s}^{\phi}\right)^{2}=\tilde{v}_{12}^{\underline{ij}}\tilde{v}_{34}^{kl}\tilde{T}_{\underline{ij}kl}.
\label{scalardoublecopy}
\end{equation}
Note that generalised dimensional reduction of the third term in \eqref{eq:numeratordouble}
gives $\left(k_{s}^{2}+\omega^{2}\right)^{2}$, but this
doesn't affect the COT or flat space limit after summing over all three
channels, so this can be discarded. After doing so, we obtain the second term in \eqref{scalardoublecopy}. The scalar-graviton numerator in \eqref{ansatzscalar} can then be written as
\begin{equation}
N_{s}^{\phi,\gamma}=\left(N_{s}^{\phi}\right)^{2}-\frac{1}{2}\left(\tilde{N}_{s}^{\phi}\right)^{2},\label{scalargravitonnum}
\end{equation}
which can in turn be expressed in terms of deformed numerators as follows:
\begin{equation}
N_{s}^{\phi,\gamma}=\frac{1}{2}\left(N_{12}^{\phi-}N_{34}^{\phi+}+N_{12}^{\phi+}N_{34}^{\phi-}\right),
\end{equation}
where
\begin{equation}
N_{12}^{\phi\pm}=N_{s}^{\phi}\pm\frac{1}{\sqrt{2}}\tilde{v}_{12}^{ij}H_{ij},\,\,\,N_{34}^{\phi\pm}=N_{s}^{\phi}\pm\frac{1}{\sqrt{2}}\tilde{v}_{34}^{ij}H_{ij}.
\end{equation}

\section{Conclusion} \label{conclusion}

In this paper we derive a compact new expression for the tree-level wavefunction of four gravitons in dS$_4$. The starting point is to write the s-channel contribution to the 4-point wavefunction for gluons as a conformal time integral, square the numerator while replacing $\nu=1/2$ Bessel functions with $\nu=3/2$ Bessel functions, and sum over all three channels. While this guess has the correct flat space limit, we need to add two more terms to the numerator in order to satisfy the COT while preserving the flat space limit. These two terms can be derived from an alternative double copy prescription, as we explain in section \ref{doublecopystruct}. The final double copy ansatz in \eqref{gravitonansatz} satisfies the flat space limit, COT, and MLT but contains spurious poles after integration. This is fixed by adding two simple correction terms: one to cancel the spurious poles and another to restore the MLT. After doing so, we obtain a four-line formula for the s-channel contribution to the graviton wavefunction in \eqref{psigamma} which agrees with the much lengthier result previously obtained in \cite{Bonifacio:2022vwa}, up to field redefinitions. As a warm-up we first carried out this procedure for massless external scalars exchanging a graviton starting from conformally coupled scalars exchanging a gluon. Remarkably, the corrections required by spurious pole cancellation and the MLT in that case directly carry over to external gravitons without any additional corrections.  

In summary, the double copy leads to significant simplifications of the 4-point graviton wavefunction in dS$_4$, although we do not yet have a systematic understanding of how it works. In particular, it would be interesting to see if there is some modification of our double copy prescription which doesn't give rise to spurious poles, and to understand the role of color/kinematics duality in our construction. In general, four-point kinematic numerators in (A)dS momentum space do not satisfy a kinematic Jacobi relation unless one performs generalised gauge transformations which modify the flat space limit \cite{Armstrong:2020woi,Albayrak:2020fyp}, but progress along these lines can be made by working in Mellin space \cite{Alday:2021odx,Mei:2023jkb}. It is worth noting that our double copy ansatz can be written in terms of asymmetric products of deformed numerators, as shown in \eqref{doublefinal} and \eqref{deformednum}, and we believe that this will be a generic feature of the double copy for non-superymmetric theories in (A)dS$_4$. Indeed, self-dual gravity in AdS$_4$ can also be derived from an asymmetric double copy with a deformed numerator\cite{sdg}, so it is conceivable that this extends beyond the self-dual sector. Another promising direction would be to translate our approach into differential notation \cite{Roehrig:2020kck,Eberhardt:2020ewh,Gomez:2021qfd,Diwakar:2021juk,Cheung:2022pdk,Herderschee:2022ntr,Lee:2022fgr,Li:2022tby}, which provides a way to uplift flat space formulae to (A)dS, but becomes complicated for spinning correlators.

Another interesting future direction would be to extend our calculations to higher points and loop-level, where expect color/kinematics duality to play an essential role. Indeed, for flat space gluon amplitudes with more than four external legs one must perform generalised gauge transformations in order to obtain numerators which obey kinematic Jacobi relations before squaring them to obtain graviton amplitudes. Evidence for color/kinematics duality at higher points was recently found for certain supersymmetric theories in AdS$_5$ \cite{Alday:2022lkk}. It may therefore be useful to generalise our method to supersymmetric theories in AdS$_5$ and compare it to recent results obtained in \cite{Alday:2021odx,Zhou:2021gnu}, as well as to supersymmetric theories in AdS$_4$ which are currently challenging to analyze using other approaches.  Since we work with general polarisations, it should be straightforward to generalise our analysis to other dimensions. Moreover the COT in \eqref{cot} and the MLT in \eqref{mlt} may naturally extend to supermomentum space, although it remains to be seen if this provides useful constraints on superwavefunctions. It would also be of  phenomenological interest to extend our construction to theories of gravity coupled to massive scalars, mimicking the flat space results in \cite{Johansson:2015oia,Plefka:2019wyg,Carrasco:2020ywq}, and to incorporate boost-breaking effects \cite{Cheung:2007st,Green:2020ebl,BBBB}. Finally, once we have a more systematic understanding of the double copy in curved background it would be of great interest to generalise the KLT bootstrap recently developed for effective field theories in flat space \cite{Chi:2021mio,Chen:2023dcx} to (A)dS$_4$, as this would complement other approaches to computing string theory corrections to AdS correlators such as conformal bootstrap methods and may even provide a useful new perspective on the UV completion of gravity in de Sitter space.  

\begin{center}
\textbf{Acknowledgements}
\end{center}
We thank Sadra Jazayeri, Silvia Nagy, Enrico Pajer, Charlotte Sleight, and David Stefanyszyn for useful discussions. CA is supported by the Royal Society via a PhD studentship. HG is supported jointly by the Science and Technology Facilities Council through a postgraduate studentship and the Cambridge Trust Vice Chancellor's Award. JM is supported by a Durham-CSC Scholarship. AL is supported by the Royal Society via a University Research Fellowship. 

\appendix

\section{Notation and Conventions}
In this Appendix we will summarise our conventions and collect various useful definitions that are used throughout the paper. When performing conformal time integrals, we Wick-rotate to Euclidean AdS$_4$ with unit radius, whose metric is given by
\begin{equation}
{\rm d}s^{2}=(1/z)^{2}({\rm d}z^{2}+{\rm d}\vec{x}^{2}),
\end{equation}
where $0<z<\infty$ is the radial coordinate and $x^i$ with $i\in\left\{ 1,2,3\right\} $ are the boundary coordinates. This is obtained from \eqref{metric} by taking $\eta \rightarrow i z$ and  dropping an overall minus sign. Moreover, we Fourier transform wavefunction coefficients to momentum space along the boundary directions and our Fourier convention is 
\eqs{
f(\vec{x})=\int \dfrac{\rd^3\vec{k}}{(2\pi)^3}{f}(\vec{k})e^{i\vec{k}\cdot\vec{x}}\equiv\int_{\vec{k}}{f}(\vec{k})e^{i\vec{k}\cdot\vec{x}}.
}

We use Greek indices, $\mu,\nu\dots$ to label the components of $4$-vectors and Latin indices from the middle of the alphabet, $i,\, j\dots$ to label the components of $3$-vectors. Latin letters from the start of the alphabet, $a,\, b\dots$ are reserved for labeling particles. The three-momenta $\vec{k}_a$ have components $k_{a}^i$ and norms $k_a =|\vec{k}_a| $. The corresponding massless four-momenta have components $k_a^{\mu} = (k_a, k_a^i)$. We define $k_{ab} = k_a+k_b$ as well as
\eqs{
 \quad k_s = | {\vec k}_1+{ \vec k}_2|, \qquad\quad k_t = | {\vec k}_1+{ \vec k}_4|, \qquad\quad k_u = | {\vec k}_1+{\vec k}_3|.
}
Using three-momentum conservation $\sum_{a=1}^4 {\vec k}_{a}=0$, these satisfy 
\eqs{ \label{NL_energy_relation}
 k_s^2+k_t^2+k_u^2=k_1^2+k_2^2+k_3^2+k_4^2.
}
We also define several combinations of these energies:
\begin{equation}\label{Es}
    E=k_{12}+k_{34},\quad E_L=k_s+k_{12},\quad E_R=k_s+k_{34},\quad \alpha=k_1-k_2,\quad\beta=k_3-k_4.
\end{equation}

We work in axial gauge where polarisation tensors only have components along boundary directions. The polarisation vectors for gluons are denoted as $\epsilon_i$ and satisfy $\epsilon_a \cdot \epsilon_a=\epsilon_a \cdot k_a=0$, where the dot denotes the product of three-vectors using the Euclidean boundary metric $\eta_{ij}$. Graviton polarisations can then be written in terms of polarisation vectors as $\epsilon_{ij}=\epsilon_i \epsilon_j$, which automatically encodes the transverse and traceless conditions. Waveunctions with external scalars can be obtained from spinning wavefunctions by the taking polarisations to satisfy $\epsilon_a \cdot \epsilon_b=1$ and $\epsilon_a \cdot k_b=0$ with $a \neq b$. The resulting scalar wavefunctions still live in the boundary of dS$_4$, so we refer to this procedure as generalised dimensional reduction. 

We use the following formulae for gluon polarisation sums, which were first defined in \cite{Baumann:2020dch}:
\eqs{
\Pi_{1,1} &= \frac{(k_1^2-k_2^2)(k_3^2-k_4^2)+k_s^2(k_u^2-k_t^2)}{k_s^4},\\
\Pi_{1,0} &= \frac{(k_1-k_2)(k_3-k_4)}{k_s^2}.
\label{psisumspin1}
}
The analogous formulae for gravitons are
\eqs{
\Pi_{2,2} &= \frac{3}{2k_s^4}(\vec{k}_1-\vec{k}_2)^i(\vec{k}_1-\vec{k}_2)^j(\pi_{il}\pi_{jm}+\pi_{im}\pi_{jl}-\pi_{ij}\pi_{lm})(\vec{k}_3-\vec{k}_4)^l(\vec{k}_3-\vec{k}_4)^m,\\
\Pi_{2,1} &= \frac{3}{2k_s^2k_{12}k_{34}}(\vec{k}_1-\vec{k}_2)^i(\vec{k}_1-\vec{k}_2)^j(\pi_{il}\hat{k}_j\hat{k}_m+\pi_{jm}\hat{k}_i\hat{k}_l+\pi_{im}\hat{k}_j\hat{k}_l+\pi_{jl}\hat{k}_i\hat{k}_m)(\vec{k}_3-\vec{k}_4)^l(\vec{k}_3-\vec{k}_4)^m,
\label{psisum}
}
where $\pi_{ij}=\eta_{ij}-\hat{k}_i\hat{k}_j$ and $\hat{k}_i = \frac{(\vec{k}_1+\vec{k}_2)_i}{k_s}$. Note that \eqref{psisumspin1} and \eqref{psisum} are defined in the $s$-channel. The equivalent expressions in the $t$- and $u$-channels can be obtained with the substitutions $2\leftrightarrow4$ and $2\leftrightarrow3$, respectively. 

\section{Integrals} \label{integrals}

In this Appendix, we will explain how to evaluate the integrals in \eqref{ansatzscalar} and \eqref{gravitonansatz}. First note that both of these formulae contain the following tensor structure:
\eqs{
2 H_{il}H_{jm}-H_{ij}H_{lm} &=\pi_{il}\pi_{jm}+\pi_{im}\pi_{jl}-\pi_{ij}\pi_{lm}\\ & +\frac{k_s^2+\omega^2}{\omega^2}\left(\pi_{il}\hat{k}_j\hat{k}_m+\pi_{im}\hat{k}_j\hat{k}_l+\pi_{jm}\hat{k}_i\hat{k}_l+\pi_{jl}\hat{k}_i\hat{k}_m\right)\\
&-\frac{k_s^2+\omega^2}{\omega^2}\left(\pi_{ij}\hat{k}_l\hat{k}_m+\pi_{lm}\hat{k}_i\hat{k}_j\right)+\left(\frac{k_s^2+\omega^2}{\omega^2}\right)^2\hat{k}_i\hat{k}_j\hat{k}_l\hat{k}_m.
\label{tensordecomp}
}
We have performed this decomposition in such a way that the first two lines encode the polarisation sums $\Pi_{2,2}$ and $\Pi_{2,1}$ in \eqref{psisum}. The final line is written in such a way as to get a convenient set of integrals. 

After performing the decomposition in \eqref{tensordecomp}, we obtain integrals of the following general form:
\eqs{
f_{A} &= \int^\infty_0\frac{d\omega\,\omega}{\omega^2+k_s^2}\int dz\,dz'(KKJ)^{3/2}_{12}(z)(KKJ)^{3/2}_{34}(z')\mathcal{I}_A\\
&= \frac{2}{\pi}\int_0^\infty\frac{d\omega\,\omega^4}{\omega^2+k_s^2}\frac{(k_{12}^2+\omega^2+2k_1k_2)}{(k_{12}^2+\omega^2)^2}\frac{(k_{34}^2+\omega^2+2k_3k_4)}{(k_{34}^2+\omega^2)^2}\mathcal{I}_A,
}
with the following set of integrands:
\eqs{
\mathcal{I}_{2,2} &=1, \quad \mathcal{I}_{2,1} =\frac{\omega^2+k_s^2}{\omega^2}, \quad \mathcal{I}_{2,0} =\left(\frac{\omega^2+k_s^2}{\omega^2}\right)^2,\\
\mathcal{I}_a &= (\omega^2+k_s^2)^2, \quad \mathcal{I}_b = (\omega^2+k_s^2).
}
The first three evaluate to
\eqs{
f_{2,2} &= \frac{2 k_1 k_2 k_3 k_4 \left(E_L E_R+E k_s\right)}{E_L^2 E^3 E_R^2}+\frac{k_1 k_2 \left(E_L k_{34}+E k_s\right)}{E_L^2 E^2 E_R}+\frac{k_3 k_4 \left(E k_s+E_R k_{12}\right)}{E_L E^2 E_R^2}+\frac{E_L E_R-k_s^2}{E_L E E_R},\\
f_{2,1} &= \frac{2 k_1 k_3 k_4 k_2}{E^3 k_{12} k_{34}}+\frac{k_1 k_2}{E^2 k_{12}}+\frac{k_3 k_4}{E^2 k_{34}}+\frac{1}{E},\\
f_{2,0} &= \frac{k_{12}k_{34}+k_s^2}{k_{12}k_{34}}f_{2,1}-\frac{k_s^2}{E}\left(\frac{k_1 k_2}{k_{12}^3 k_{34}}+\frac{2 k_1 k_3 k_4 k_2}{k_{12}^3 k_{34}^3}+\frac{k_3 k_4}{k_{12} k_{34}^3}\right),
\label{integrals1}
}
where $E ,\, E_L$, and $E_R$ are defined in \eqref{Es}. The last two integrals are divergent: 
\eqs{
f_a &= \frac{2}{\pi}\left(\frac{\Lambda^3}{3}-\Lambda(k_{12}^2+k_{34}^2-k_s^2+2(k_1k_2+k_3k_4))\right)+\mathrm{finite},\\
f_b &= \frac{2}{\pi}\Lambda + \mathrm{finite},
}
where $\Lambda$ is a cut-off on the $\omega$ integral. On the other hand, the divergent pieces are analytic in at least two of the momenta and therefore correspond to boundary contact terms. Moreover they become imaginary after analytically continuing back to de Sitter so won't contribute to the in-in correlator. Dropping these divergences then gives
\eqs{
f_a &= \left(k_{12}k_{34}+k_s^2\right)f_b+\frac{1}{E}\left(2k_1k_2k_3k_4-k_1k_2(2E^2+k_{12}^2)-k_3k_4(2E^2+k_{34}^2)-2k_{12}k_{34}E^2+E^4\right),\\
f_b &= \left(\frac{2k_1k_2k_3k_4}{E^3}+k_1k_2\frac{k_{34}+E}{E^2}+k_3k_4\frac{k_{12}+E}{E^2}+\frac{k_{12}k_{34}-E^2}{E}\right).
\label{integrals2}
}

\bibliography{DCrefs}
\bibliographystyle{JHEP}
  
\end{document}